\documentclass[useAMS,usenatbib]{mn2e}

\usepackage{graphicx} 
\usepackage{txfonts}
\usepackage{lscape}
\usepackage{mathrsfs}
\usepackage{nicefrac}
\usepackage[german]{babel}

\title[Measuring the total and baryonic mass profiles of the very massive CASSOWARY 31 strong lens]{Measuring the total and baryonic mass profiles of the very massive CASSOWARY 31 strong lens. A fossil system at $z \simeq 0.7$?\thanks{Based on data from the X-shooter GTO observations collected at the European Southern Observatory VLT/Kuyuen telescope, Paranal, Chile, under the programme ID 089.A-0222(A).}}

\author[C. Grillo et al.]{C. Grillo$^{1}$\thanks{E-mail: grillo@dark-cosmology.dk}, L. Christensen$^{1}$, A. Gallazzi$^{1,2}$ and J. Rasmussen$^{1,3}$ \\
$^{1}$Dark Cosmology Centre, Niels Bohr Institute, University of Copenhagen, Juliane Maries Vej 30, 2100 Copenhagen, Denmark \\
$^{2}$INAF - Osservatorio Astrofisico di Arcetri, Largo Enrico Fermi 5, 50125 Firenze, Italy \\
$^{3}$Department of Physics, Technical University of Denmark, Ris\o\ Campus, Frederiksborgvej 399, 4000 Roskilde, Denmark}

\begin{document}

\date{Accepted. Received; in original form}

\pagerange{\pageref{firstpage}--\pageref{lastpage}} \pubyear{}

\maketitle

\label{firstpage}

\begin{abstract}
We investigate the total and baryonic mass distributions in deflector
number 31 of the Cambridge And Sloan Survey Of Wide ARcs in the skY
(CASSOWARY). We confirm spectroscopically a four-image lensing system
at redshift 1.4870 with VLT/X-shooter observations. The lensed images
are distributed around a bright early-type galaxy at redshift 0.683,
surrounded by several smaller galaxies at similar photometric
redshifts. We use available optical and X-ray data to constrain the
deflector total, stellar, and hot gas mass through, respectively,
strong lensing, stellar population analysis, and plasma modelling. We
derive a total mass projected within the Einstein radius $R_{\mathrm{Ein}}=70$ kpc of $(40 \pm 1)\times 10^{12}$ M$_{\odot}$, 
and a central logarithmic slope of $-1.7 \pm 0.2$ for the total mass density. Despite a very high stellar mass and velocity
dispersion of the central galaxy of $(3 \pm 1) \times 10^{12}$~M$_{\odot}$ and $(450 \pm 80$) km s$^{-1}$, respectively, the cumulative stellar-to-total mass profile of the
deflector implies a remarkably low stellar mass fraction of 20\% (3\%--6\%) in projection within the central galaxy effective radius $R_{e}=25$~kpc ($R=100$~kpc). We also find that the CSWA 31 deflector has properties suggesting it to be among the most distant and massive fossil systems studied so far. The unusually strong central dark matter dominance and the possible fossil nature of this system renders it an interesting target for detailed tests of cosmological models and structure formation scenarios.

\end{abstract}


\section{Introduction}

Strong gravitational lensing has become a powerful tool that can be used to address a number of problems in modern cosmology and galaxy evolution (e.g., \citealt{bar10,tre10b}). Given the fact that the gravitational deflection of light is determined only by the gravitational field through which light propagates, the gravitational lensing effect is independent of the nature of the matter and of its state. This implies that lensing is able to measure both dark and baryonic matter, in equilibrium or far out of it. From the image configuration of a lensing system the total mass of the lens, within a cylinder with a diameter of the average image separation and centred on the lens, can be estimated very accurately. Mass measurements with a precision of a few percent can be achieved by means of detailed lens models in multiple image systems (e.g., \citealt{gri10c,zit12}). These are by far the most precise mass determinations in extragalactic astronomy. Strong gravitational lensing has already yielded groundbreaking results detecting low-mass dark-matter sub-haloes without visible stars in a few lens galaxies (e.g., \citealt{veg10,veg12}). Thanks to the combination of strong gravitational lensing with stellar dynamics and/or population synthesis models (e.g., \citealt{gri12,new12,bar12,son12}), the prospects for making a step forward in our understanding of the nature of dark matter are very promising. The combination of these different mass diagnostics has also proved to be extremely successful in the exploration of several other astrophysical and cosmological topics, such as the study of the inner projected dark over total mass fractions (e.g., \citealt{gri09,aug09,eic12}), stellar initial mass function (e.g., \citealt{gri09,tre10,spi11}), and tilt of the Fundamental Plane (e.g., \citealt{gri09,gri10b}) of massive early-type galaxies and the investigation of the values of the cosmological parameters (e.g., \citealt{gri08c,schw10,suy10,suy13}).

In this paper, we study the total and baryonic mass distributions of the deflector number 31 (CSWA 31) of the Cambridge And Sloan Survey Of Wide ARcs in the skY (CASSOWARY) survey (\citealt{bel09}). This complex strong lensing system was first analysed by \citet{brewer11}, using a diffusive nested sampling technique for the lens modelling and focussing on the reconstruction of the unlensed source profiles and deflector total mass distribution. The remarkably large Einstein radius of 70 kpc of this system makes it an ideal candidate to investigate the still relatively unexplored amount and distribution of dark matter in the central regions of galaxy groups (e.g., \citealt{lim09,gri11}). We present here a significant improvement in the characterisation of the internal structure of this deflector by including new spectroscopic data. In particular, by measuring the redshift of the main lensing system, we obtain accurate estimates of the projected stellar-to-total mass profile in the deflector inner regions. 

The manuscript is organised as follows. In Sect. 2, we present our spectroscopic observations obtained with VLT/X-shooter. In Sect. 3, we perform a strong lensing analysis to determine the deflector total mass profile. In Sect. 4, we estimate the deflector baryonic mass profile, both in terms of luminous (stellar) and hot gas components. In Sect. 5, we discuss why the CSWA 31 deflector is an unconventional strong lens and a potential fossil system. In Sect. 6, we summarise the main results. Throughout this work we assume $H_{0}=70$ km s$^{-1}$ Mpc$^{-1}$, $\Omega_{m}=0.3$, and $\Omega_{\Lambda}=0.7$. In this model, 1\arcsec$\ $ corresponds to a linear size of 7.08 kpc at the deflector redshift of $z_{\mathrm{sp}}=0.683$. 

\section{Observations}

\begin{figure}
  \centering
  \includegraphics[width=0.48\textwidth]{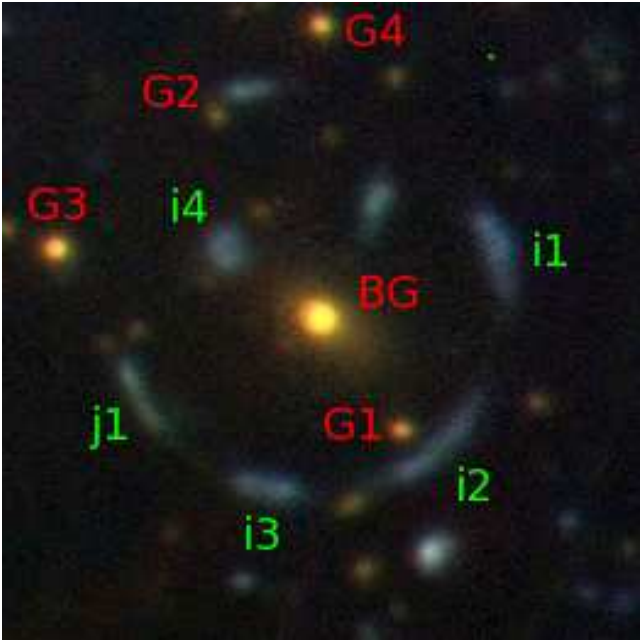}
  \caption{Colour-composite image (38 $\times$ 38 \arcsec, i.e., 269 $\times$ 269 kpc on the deflector plane) of the strong lensing system CASSOWARY 31 obtained by combining the Gemini/GMOS $g$, $r$, and $i$ bands. }
  \label{fig1}
\end{figure}

\begin{table}
\centering
\caption{The Gemini and SDSS data of the lens galaxies.}
\label{tab1}
\begin{tabular}{cccccc} 
\hline\hline \noalign{\smallskip}
Object & R.A. & Dec. & $z_{\mathrm{ph}}$ & $d\,^{a}$ & $\theta_{e,i}$ \\
& (J2000) & (J2000) & & (\arcsec) & (\arcsec) \\
\noalign{\smallskip} \hline \noalign{\smallskip}
BG & 09:21:25.74 & 18:10:17.3 & $0.64 \pm 0.04\,^{b}$ & 0.0 & 3.54 \\
G1 & 09:21:25.89 & 18:10:09.2 & $0.71 \pm 0.09$ & 8.2 & 0.29 \\
G2 & 09:21:25.37 & 18:10:30.5 & $0.78 \pm 0.22$ & 13.7 & 2.36 \\
G3 & 09:21:26.24 & 18:10:32.1 & $0.78 \pm 0.05$ & 16.3 & 1.13 \\
G4 & 09:21:24.82 & 18:10:28.9 & $0.73 \pm 0.09$ & 17.5 & 0.48 \\
\noalign{\smallskip} \hline
\end{tabular}
\begin{list}{}{}
\item[$^a$]Relative to the BG luminosity centre.
\item[$^b$]Spectroscopic redshift $z_{\mathrm{sp}}=0.683$.
\end{list}
\end{table}

\begin{figure}
  \centering
  \includegraphics[width=0.46\textwidth]{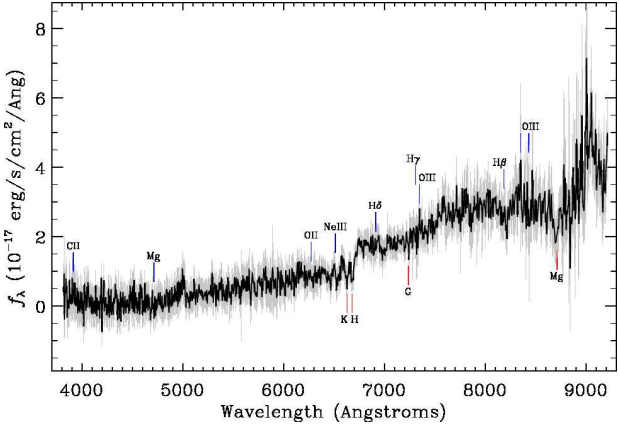}
  \caption{SDSS spectrum of the BG measured with a 3\arcsec-diameter fibre. Several emission and absorption lines are identified and used to estimate a redshift value $z_{\mathrm{sp}}$ of 0.683.}
  \label{fig4}
\end{figure}

The CSWA 31 strong lensing system was observed with GMOS on Gemini South on February 21, 2009 (programme ID GS-2009A-Q-64; \citealt{brewer11}), and the data
were retrieved through the Gemini science archive. The total integration times were 20 minutes per filter in the $g$, $r$ and $i$-band split into 4, 4 and 8 exposures, respectively, with offsets between each exposure. The data were reduced and combined using standard methods, and a colour image of the system is presented in Fig. \ref{fig1}. In this figure, we label the five innermost and most massive galaxies that we associate with the deflector (i.e., BG and G1-G4) and the lensed objects (i.e., i1-i4 and j1) that will be discussed in the following. The brightest galaxy (BG) coordinates from the Gemini data and spectrum from the SDSS are shown, respectively, in Table \ref{tab1} and Fig. \ref{fig4}. In the same table, we also list the Gemini coordinates of the four galaxies (G1-G4), detected in the SDSS, with photometric redshift values $z_{\mathrm{ph}}$ that are consistent, given the errors, with the spectroscopic redshift value $z_{\mathrm{sp}}=0.683$ of the BG, their projected distances $d$ from the luminosity centre of the BG and the values of their $i$-band effective angles $\theta_{e,i}$, estimated by fitting the SDSS luminosity profiles.

To determine spectroscopic redshifts of the multiply imaged sources we
obtained slit spectra with VLT/X-shooter \citep{vernet11} on April 26,
2012 (programme ID 089.A-0222(A)) using several different pointings.
Despite poor observing conditions with variable seeing, high wind and
humidity, we were able to secure spectra of three of the lensed images (i1, i3, and j1). In a first observational configuration, we nodded between images i1 and j1 with exposures of 900 seconds on each position in an ABBA sequence. The objects were placed on the X-shooter 11\arcsec-long slit in such a way that the alternate position could be used for the sky subtraction in the NIR, while in the UVB and VIS the background sky was measured in the same slit. The chosen slit widths were 1.0/0.9/0.9 arcsec in the
UVB, VIS and NIR arms, respectively. In a second configuration, we obtained a single integration of 900 seconds on image i3 and a faint blue galaxy 16.5\arcsec\, away from the centre of the BG. For the sky subtraction, we used a separate integration of 900 seconds of a nearby blank sky region. Due to a higher seeing the slit widths were increased to 1.3/1.2/1.2 arcsec in the three arms.

\begin{figure}
  \centering
  \includegraphics[width=0.50\textwidth]{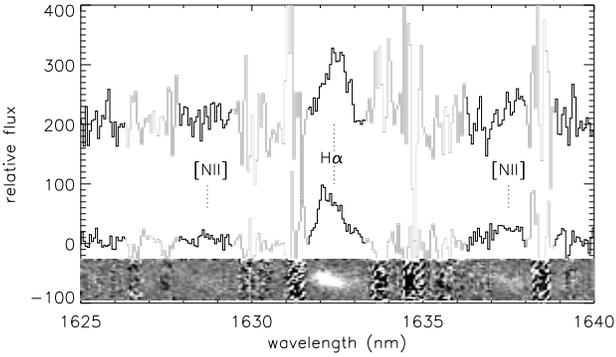}
  \caption{From the bottom, two-dimensional and one-dimensional spectra of image i1 around the redshifted H$\alpha$ and [NII] lines. The one-dimensional spectrum of image i3 is also shown for comparison offset along the y-axis. The residuals from strong sky emission lines are masked.}
  \label{fig2}
\end{figure}

The data were reduced with the ESO pipeline version 1.3.7
\citep{modigliani10}. The entire wavelength range from 300 to 2500 nm
was examined for the presence of emission lines to measure the
redshift of the sources. Typically, the continuum emission was very
faint and could only be recognised in binned versions of the 2D
spectra \citep[see][]{christensen12}. In the spectrum of images i1 and i3, we detect a single bright line
in the NIR. This emission line is interpreted as H$\alpha$ at
$z=1.4870\pm0.0003$. Weaker lines corresponding to [NII]
$\lambda\lambda$6548,6586 can also be seen, as illustrated in Fig. \ref{fig2} in the apposite
wavelength range of the observed two-dimensional spectrum.
H$\beta$ and [OIII] are detected at the 5- and 4-sigma significance level in the extracted 1D spectra, respectively, at the wavelengths corresponding to the previously cited redshift. The 1D spectra of the two exposures of the different images shown in Fig. \ref{fig2} prove that the two objects are lensed multiple images of the same background source. The spectrum of image j1 reveals a single, bright emission line at
18859.4 {\AA}. No other emission lines are detected in the entire
wavelength range. By binning the UVB spectrum by a factor of 500 in
the spectral direction, we detect the
continuum emission at least down to approximately 4000 {\AA}, suggesting a
redshift $z<2.3$. The most likely interpretation is therefore that the
single emission line is H$\alpha$ at $z=1.8737\pm0.0003$.

\section{Total mass measurements through strong lensing modelling}

\begin{table}
\centering
\caption{The median values and 68\% CL intervals (relative errors are shown in square brackets) of the lens parameters from bootstrapping analyses on 3000 resampled positions of the multiple images. }
\label{tab2}
\begin{tabular}{ccccc} 
\hline\hline \noalign{\smallskip}
Model & $b\,^{a}$ & $q$ & $\theta_{q}$ & $\gamma_{\mathrm{T}}$ \\
& (\arcsec) & & (deg) & \\
\noalign{\smallskip} \hline \noalign{\smallskip}
SIE & 9.0$^{+0.2}_{-0.2}\,[_{-2\%}^{+2\%}]$ & 0.53$^{+0.05}_{-0.04}\,[_{-8\%}^{+9\%}]$ & $-$46$^{+1}_{-1}\,[_{-2\%}^{+2\%}]$ & \\
EPL & 14$^{+6}_{-3}\,[_{-21\%}^{+43\%}]$ & 0.73$^{+0.08}_{-0.10}\,[_{-14\%}^{+11\%}]$ & $-$45$^{+1}_{-1}\,[_{-2\%}^{+2\%}]$ & $-$1.7$^{+0.2}_{-0.2}\,[_{-12\%}^{+12\%}]$ \\
\noalign{\smallskip} \hline
\end{tabular}
\begin{list}{}{}
\item[$^a$]Note that Gravlens provides the value of the lens strength $b$ multiplied by a function $f(\cdot)$ of the minor-to-major axis ratio, $q$, and of the exponent of the three-dimensional density distribution, $\gamma_{\mathrm{T}}$ (see \citealt{kee01a,kee01b}).
\end{list}
\end{table}

In this section we focus on the measurement of the projected total mass profile of the deflector by modelling the observed multiple images.

We use the public code Gravlens\footnote{http://redfive.rutgers.edu/$\sim$keeton/gravlens/} (\citealt{kee01a}) to perform our analysis. We describe the total mass distribution of the deflector in terms of a one-component model, chosen to be either a singular isothermal ellipsoid (SIE) or an elliptical power-law (EPL) profile (for further details on these mass profiles, see \citealt{kee01b}). We fix the total mass centre of the deflector to the luminosity centre of the BG. Since images i1, i2, i3, and i4 have very similar colours (see Fig. \ref{fig1}), we assume that i2 and i4 are additional multiple images of the source at redshift 1.4870 (see Sect. 2) that is lensed into i1 and i3. We approximate the four multiple images i1-i4 to point-like objects and conduct a standard strong lensing chi-square optimisation on the model parameters (e.g., \citealt{gri08b,gri10c,gri11}) with three and two degrees of freedom for the SIE and EPL models, respectively. We adopt a positional uncertainty of 0.3\arcsec\, (corresponding to two image pixels) on all image positions and repeat the strong lensing analysis starting from the measurements of the luminosity peaks of the multiple images in each of the three ($g$, $r$, and $i$) available Gemini images. We obtain best-fitting (minimum) $\chi^{2}$ values of 10.8 and 9.0, respectively, for the SIE and EPL models. This translates into average differences between observed and model-predicted positions of approximately three pixels per image. Given the approximation to point-like objects centred on their luminosity peaks for the extended images and the assumption of a single total mass component fixed on the deflector luminosity centre, we conclude that our simplified modelling choices can reproduce well the observations and do not need to be further refined, having as main goal here the measurement of the projected total mass profile of the deflector. In particular, we decide to avoid introducing an unnecessary, at this stage, external shear term for the following reasons: first, as discussed, the strong lensing configuration can be reconstructed in a satisfactory way by a single total mass component; second, the limited number of degrees of freedom available from the use of only four multiple images would not allow to disentangle the contribution of a possible external shear term from that of the main deflector mass ellipticity; third, an external shear term would not affect the projected total mass estimates, that are the most important quantities for the deflector mass decomposition presented in Sect. 5.1. 

To estimate the statistical errors on the projected total mass profile, we rerun the chi-square optimisation on 3000 resampled positions of the multiple images. For these last quantities, we assume normal distributions with mean and standard deviation values equal to the observed positions and adopted positional uncertainties, respectively. We calculate the median values and the 68\% (95\%) confidence level (CL) intervals of the lens parameters by removing from the 3000 optimised values the 480 (75) lower and higher values and show the results in Table \ref{tab2}. We remark that the values of the total mass minor-to-major axis ratio, $q$, and of the position angle of the major axis, $\theta_{q}$, are in line with those derived from a fit to the extended light distribution in the $i$-band of the BG ($q_{BG} = 0.47 \pm 0.01$ and $\theta_{q,BG} = -41 \pm 1$). We notice that the reconstructed total mass distribution does not show an unusually large elongation, further supporting the adequacy of a one-component model (without external shear) to represent the deflector total mass. We also find that the total mass has a slightly less elliptical distribution than the light (of the BG), as already observed in several previous strong lensing studies (e.g., \citealt{bar11,son12}). For the more general EPL model, we estimate a total mass value $M_{\mathrm{T}}$ of $(4.0 \pm 0.1)\times 10^{13}$ M$_{\odot}$ projected within a cylinder of radius of 70 kpc, which is the approximate value of the Einstein radius $R_{\mathrm{Ein}}$ of this strong lensing system. We plot in the top panel of Fig. \ref{fig3} the median, 68\% and 95\% CL values of the cumulative total projected mass of the deflector (we have checked that the differences between the reconstructed SIE and EPL total mass profiles do not have a significant impact on our overall conclusions). From the results of the same resampling analysis, we also estimate a median value of $-1.7$ and a 68\% CL uncertainty of 0.2 for the three-dimensional, radially averaged, logarithmic slope of the total density profile $\gamma_{\mathrm{T}}= d \log [\rho_{\mathrm{T}}(r)] / d \log (r)$ of the deflector (see Table \ref{tab2}), without assuming any prior on this parameter. The inspection of the $\chi^{2}$ distribution as a function of the parameter $\gamma_{\mathrm{T}}$ analogously reveals a rather sharply peaked global minimum at approximately $-1.7$. Furthermore, we will show in Sect. 5.1 that this measurement is consistent, given the errors, with the outcome of an initial lensing$+$dynamics study.

We have explicitly tested through simulations that our resampling technique with point-like images allows to recover for the deflector not only the total projected mass within the Einstein radius, but also the slope of the total mass density profile with sufficient accuracy (i.e., the true and reconstructed values agree well within the uncertainties). This is possible because the four multiple images observed in CSWA 31 are not all located at the same projected distance from the deflector centre (see Fig. 1). The position of image i4, significantly closer than the other images to the adopted mass centre of the deflector (see also Fig. 4), is particularly useful to estimate the slope of the deflector total mass density. In detail, we have simulated three strong lensing configurations very similar to that of the quad in the CSWA 31 system and measured the values of the deflector total density slopes with the same method used above. We have chosen three different values, i.e. $-1.7$, $-2.0$, and $-2.3$, of the density slope $\gamma_{\mathrm{T}}$ of the elliptical power-law profile adopted to describe the total mass distribution of the deflector and left all the other parameters of the deflector unchanged. The source has been fixed approximately at the reconstructed position of the source for the quad of the CSWA 31 system. The redshifts of deflector and source have been chosen equal to those of the CSWA 31 system. For each of the three models we have used the ray-tracing equation to obtain the positions of the corresponding four multiple images and estimated the approximate Einstein radii of the systems as the mean distances of the images from the deflector centre. Then, for each model, we have resampled 1000 times the positions of the four images assuming normal distributions with mean and standard deviation values equal to the positions previously derived and 0.3\arcsec$\,$ and minimised the chi-square varying the deflector parameters and starting from the approximate Einstein radius values and an initial value of $\gamma_{\mathrm{T}}$ of $-2$. We have used the optimised deflector parameters to calculate the median values and the 68\% CL intervals of $\gamma_{\mathrm{T}}$. We can always find the true values of $\gamma_{\mathrm{T}}$ within our measured 68\% CL intervals and the median values of the estimated $\gamma_{\mathrm{T}}$ are relatively close to the corresponding true values, despite some modest sistematic bias towards larger values. The origin of this small overestimation of the median values might be associated to the particular geometrical configuration of the multiple images chosen in our simulations to mimic the CSWA~31 system or to the slightly asymmetric shape of the $\chi^2$ distribution as a function of $\gamma_{\mathrm{T}}$ (less steep on the side of larger values of $\gamma_{\mathrm{T}}$). Considering that we never use in our analysis a single value of $\gamma_{\mathrm{T}}$, but always the 68\% and 95\% CL intervals for this parameter, we are confident that the measurements presented in this work for $\gamma_{\mathrm{T}}$ in CSWA 31 are robust and not significantly biased. As already mentioned, the remarkably good consistency of the lensing-only and lensing+dynamics (see Sect. 5.1) results for the value of $\gamma_{\mathrm{T}}$ in CSWA 31 reinforces this last point.

We notice that j1 seems to have colours similar to those of the radially extended image located close in projection to the BG, between i1 and i4. If j1 and this object were multiple images of the same source at $z = 1.8737$ (see Sect. 2), according to our lensing models we would expect to observe two additional images, like in the i1-i4 system. If instead a counter image of j1 were the tangentially elongated object angularly adjacent to G2 (as assumed in \citealt{brewer11}), we would also predict the presence of other multiple images near i2. We do not detect any clear four-image system for j1 in the actual data. Because of the not obvious geometrical configuration of the lensing system to which j1 belongs, we decide not to include this last system in our models until new photometric and/or spectroscopic data become available. 

\section{Baryonic mass measurements}

In this section we concentrate on the estimate of the baryonic (stellar and hot gas) matter component associated with the deflector. 

\subsection{Composite stellar population modelling}

Optical colors are known to correlate well with the galaxy effective mass-to-light ratio \citep[e.g.,][]{BdJ01,zib09}. 
In order to derive constraints on the stellar mass content of the deflector we interpret the SDSS optical photometry
of the five members listed in Table \ref{tab1} by means of a Monte Carlo library of star formation histories (SFH) and dust
attenuation adopting a Bayesian approach, as in \citet{gal05}. The SFHs are modelled with exponentially declining laws on top of which
random bursts\footnote{The formation time is required to be younger than the age of the Universe at the redshift
of the BG, $z=0.683$. The star formation timescale can vary between 0 and 1~Gyr$^{-1}$. The fraction of stellar
mass produced in a burst is logarithmically distributed between 0.03 and 4, and the duration of the burst can vary
between $3\times10^7$ and $3\times10^8$ yr. The models span a range of metallicities from 0.1 to 2 times solar.}
can occur with a probability such that 50\% of the models in the library experience a burst in the last 2~Gyr
\citep[see also][]{Salim05}. Attenuation by dust follows the \cite{CF00} model and is parametrized by the total
effective optical depth $\tau_V$ experienced by young stars in their birth clouds (which can vary between 0 and 6)
and the fraction $\mu$ contributed by the interstellar medium (which can vary between 0.1 and 1). Our library of models is constructed adopting a \cite{chabrier03} IMF, however for the following analysis we decide to rescale the derived stellar masses to a \cite{sal55} IMF, assuming a conversion factor of 1.8. This is motivated by the results of several recent analyses (e.g., \citealt{gri09}; \citealt{van10}; \citealt{gri10b}; \citealt{tre10}; \citealt{con12}), according to which massive early-type galaxies (as those studied here) are better represented by a bottom-heavy stellar IMF.

Because of the low and very uncertain flux level in the $u$ and $g$ bands, resulting in colours inconsistent with
the other colours at redder wavelengths predicted by the models, we decide to discard the observed magnitudes in these
two bands and use only the observed $r, i, z$ SDSS model magnitudes, corrected for foreground galactic reddening (see also \citealt{gri09,gri11}).
These are compared with those computed on all the model spectra in the library. The stellar mass of each model is
obtained by multiplying the mass currently present in stars (i.e. not including the fraction returned to the ISM
by evolved stars) with the scaling factor to the observed fluxes. We then construct the probability density
function of stellar mass, the median of which is our fiducial $M_\ast$ estimate while the $16^{th}$ and $84^{th}$
percentiles define the 1-$\sigma$ confidence interval. The results are summarized in Table~\ref{tab_mstar}. 

We find a very high stellar mass of approximately $\rm 3\times10^{12}\,M_\odot$ for the BG with a formal uncertainty of
approximately $40$\%, and stellar masses around or above $\rm 10^{11}\,M_\odot$ for other three members of the system, albeit
with very large uncertainties. We explored the systematic effects on the derived stellar masses due to our
assumptions for the model library. In particular we repeated the fit 1) using only metallicities around solar, 2)
excluding bursty SFHs or 3) assuming very low levels of dust attenuation. In Table~\ref{tab_mstar} we provide
the difference between our default fit and the modified one as an estimate of the possible systematic
uncertainty.
We find that the assumption on
metallicity does not affect the results, as the red colors of the galaxies require a high metallicity. Removing
the younger component introduced by the bursts typically provides higher stellar masses by up to 0.1~dex. Assuming
negligible dust attenuation results in lower stellar masses by 0.1~dex for the BG and more than 0.2~dex for the
other members. We note however that models without dust hardly reach the red $r-i$, $i-z$ observed colours
providing thus a worse fit to the data.

\begin{figure}
  \centering
  \includegraphics[width=0.46\textwidth]{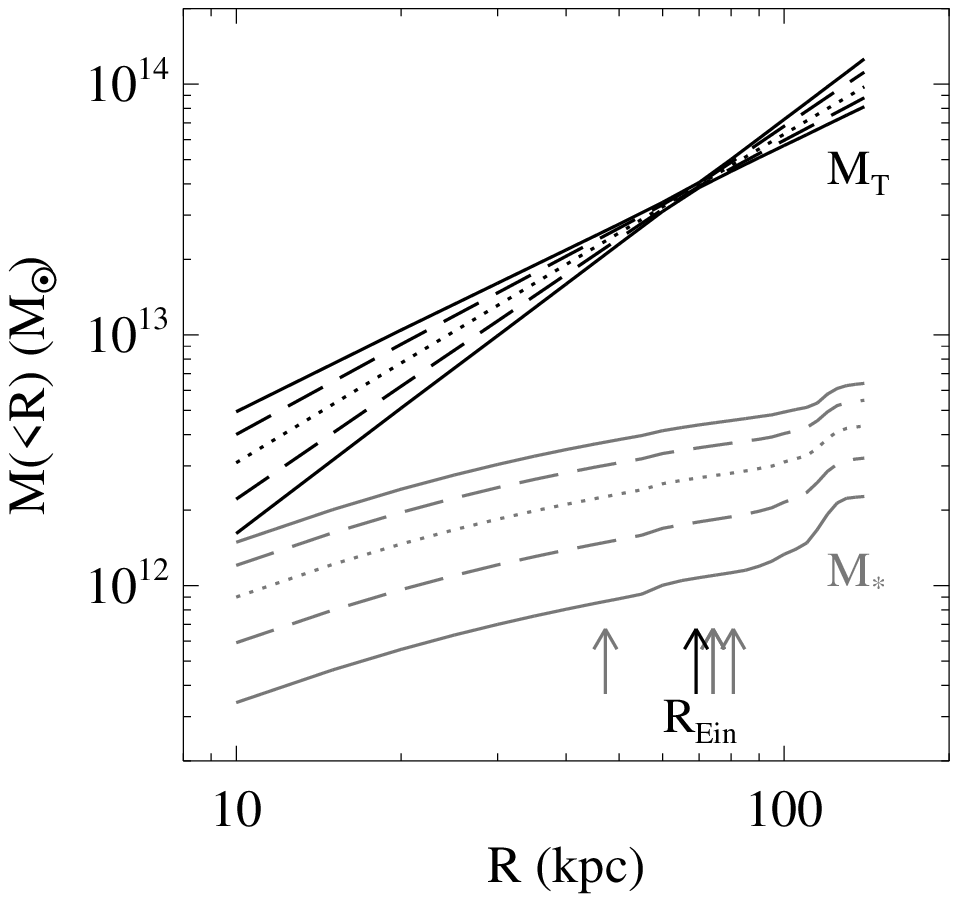}
  \includegraphics[width=0.46\textwidth]{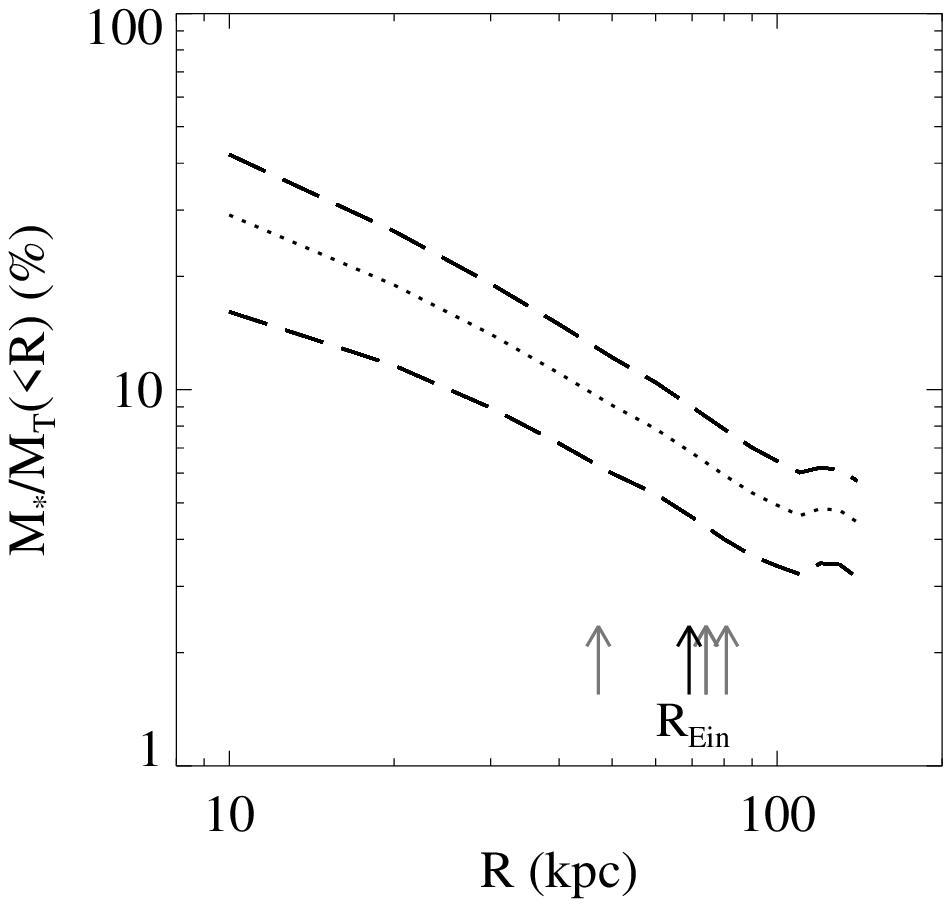}
  \caption{Cumulative stellar, $M_*$, and total (derived for the EPL model), $M_{\mathrm{T}}$, masses (on the top) and their ratio, $M_*/M_{\mathrm{T}}$, (on the bottom) projected within cylinders of radii $R$. The dotted lines represent the median values; the dashed and solid lines show the 68\% and 95\% CL intervals, respectively. The four grey arrows (two of them overlap) locate the projected distances of the observed multiple images i1-i4 from the BG luminosity centre; the black arrow indicates the Einstein radius $R_{\mathrm{Ein}}$ of the strong lensing system.}
  \label{fig3}
\end{figure}

Finally, we mention that according to the recent study by \citet{cap12} the most massive ellipticals might have a maximal IMF, resulting in stellar mass-to-light ratios approximately 1.5 times higher than those obtained from a Salpeter IMF. In our system, this difference would apply at most to the BG galaxy. The other, smaller, galaxies should have a lighter IMF, thus bringing their estimated mass to lower values than for a Salpeter IMF. The final effect on the stellar mass profile shown in Fig. \ref{fig3} would be a small shift of 0.17 dex higher than what we find now (the contribution of the other galaxies would be even smaller). We notice that this possible shift can be accommodated by our present 95\% CL interval plotted in Fig. \ref{fig3}. Therefore, this would not change significantly our final results.

\begin{table}
\centering
\caption{Stellar mass estimates from $r,i,z$ SDSS model magnitudes for the BG and the other four galaxies G1-G4, assuming a Salpeter IMF. The last three columns give an estimate of the possible systematic uncertainties associated with different assumptions
on metallicity, burstiness of the SFHs and dust attenuation, respectively.}\label{tab_mstar}
\begin{tabular}{|l|c|r|r|r|}
\hline\hline \noalign{\smallskip}
Object & log$(M_*/$M$_\odot)$ & $\rm \Delta$$_{Z=Z_\odot}$ & $\rm \Delta_{no bursts}$ & $\rm \Delta_{no dust}$\\
\hline
\noalign{\smallskip}
BG & $12.48^{+0.16}_{-0.13}$ & $-0.02$ & $-0.09$ & $+0.10$ \\
G1 & $10.84^{+0.34}_{-0.38}$ & $-0.02$ & $+0.10$ & $+0.20$ \\ 
G2 & $11.27^{+0.35}_{-0.37}$ & $ 0.00$ & $+0.04$ & $+0.24$ \\
G3 & $11.89^{+0.27}_{-0.22}$ & $ 0.00$ & $-0.10$ & $+0.27$ \\ 
G4 & $11.34^{+0.33}_{-0.28}$ & $+0.01$ & $-0.06$ & $+0.24$ \\ 
\noalign{\smallskip}
\hline
\end{tabular}
\end{table}

\subsection{X-ray emission modelling}

To estimate the mass of any hot intracluster medium (ICM) within the
central regions, we searched for known X-ray sources close to the
BG. X-ray data for this field are limited to those of the {\em ROSAT}
All-Sky Survey (RASS), in which the nearest detected source
lies $(26\pm 13)\arcsec$ away from the BG position. With only 7~net
counts, the source is listed as point-like but with a poorly
constrained spatial extent. The X-ray error circle includes none of
the CSWA 31 candidate members but does overlap with an SDSS galaxy at
unknown redshift. If representing extended cluster emission with a
spectrum described by a thermal plasma (APEC) model, the quoted RASS
hardness ratios of the source would imply a gas temperature $T>2$~keV
for any subsolar metallicity. However, the hardness ratios can be
equally well reproduced by a (potentially intrinsically obscured)
power-law spectrum with photon index $\Gamma < 2$, characteristic of
AGN. Furthermore, the spatial offset from the BG/deflector centre of
$(26\pm 13)\arcsec$ would correspond to a projected distance of $(184\pm
92)$~kpc at $z=0.683$, implying an offset
between the dark matter and X-ray peaks comparable to that of the
famous Bullet cluster \citep{clow06}. This large offset, combined with
the presence of a galaxy within the X-ray error circle, the fact that
a power-law nature of the X-ray spectrum cannot be excluded, and that
the emission appears point-like (albeit with low robustness), lead us
to conclude that this source is unlikely to represent ICM emission
from CSWA 31.

The region covered by Fig.~\ref{fig3} would appear point-like in
RASS, so any ICM signal within this aperture must have a count rate
below that of the nearby source discussed above. This constraint can
be translated into a limit on the corresponding gas mass for a given
ICM temperature. To this end, we first assume that the total mass
profile in Fig.~\ref{fig3}, when corrected for the stellar
contribution, can be
described by an NFW profile \citep*{nava97} with concentration
parameter, $c_{200} < 10$. Using the inferred $M_{\rm T}$ and $M_{\rm
  *}$ within $R=150$~kpc as constraints, this would suggest a total
dark matter mass of $M_{\rm DM} > 6 \times 10^{14}$~M$_\odot$ within
the resulting $R_{500} > 1.0$~Mpc.
For this mass limit, the observed $M$--$T_{\rm X}$ relation for X-ray
selected clusters, corrected for self-similar redshift evolution
\citep{vikh06}, would suggest $T \ga 6$~keV. Using these constraints,
and assuming any subsolar ICM metallicity, a {\sc
  pimms}\footnote{http://heasarc.gsfc.nasa.gov/Tools/w3pimms.html}
calculation shows that the absence of detectable ICM emission in the
RASS data implies $L_{\rm X} < 4 \times 10^{44}$~erg~s$^{-1}$
(0.5--2~keV rest-frame) and $M_{\rm gas} < 3 \times 10^{12}$~M$_\odot$
within $R=150$~kpc. This indicates that within this aperture only a minor fraction of our estimated $M_{\rm T}$ could be in the form of hot gas and that the stellar
component is the main baryonic component.

\section{Discussion}

\subsection{CSWA~31 as an unusual deflector}

We estimate here the variation in the deflector stellar-to-total mass fraction $M_{*}/M_{\mathrm{T}}$ projected within cylinders of increasing aperture values $R$ from the BG centre. We use the total mass measurements obtained in Sect. 3 for the EPL model (the different adoption of the SIE model would not change appreciably the following conclusions) and the galaxy $i$-band effective angles $\theta_{e,i}$ and stellar masses $M_{*}$ that are listed, respectively, in Tables \ref{tab1} and \ref{tab_mstar}. We build pixellated grids of projected stellar mass assuming that the latter is well traced by the light and that the de Vaucouleurs profiles with effective angles $\theta_{e,i}$ for BG and G1-G4 are good representations of their light distributions. We bootstrap the galaxy stellar mass values according to the statistical uncertainties (see Sect. 4.1). We sum the contribution of each galaxy, calculate the cumulative projected stellar mass profile (see Fig. \ref{fig3}, on the top), and divide it by the cumulative projected total mass profile. We plot the radial dependence of the ratio $M_{*}/M_{\mathrm{T}}(<R)$ and its 68\% CL errors on the bottom of Fig. \ref{fig3}. As expected, we observe that the relative error on the total mass estimate increases moving away from the Einstein radius of the system and that the stellar-to-total mass fraction decreases rapidly towards the outer regions of the deflector. This translates into a significant amount of dark-matter present already at relatively small distances from the BG centre, given the upper limit on the hot gas mass provided in the previous section. More quantitatively, at a projected distance of 25 kpc, i.e. at the value of the BG effective radius, only approximately 20\% of the total mass is in the form of stellar mass. In the interval $R=100$--150 kpc, the 68\% CL values of $M_{*}/M_{\mathrm{T}}(<R)$ do not vary significantly and range between 3\% and 6\%. Furthermore, if we neglect the hot gas contribution between 10 and 50 kpc and estimate here the dark matter profile of the deflector as the difference between the total and stellar mass components, we find that, assuming a simplified spherical power-law profile, the three-dimensional inner logarithmic slope of the dark matter halo $\gamma_{\mathrm{DM}}= d \log [\rho_{\mathrm{DM}}(r)] / d \log (r)$ is equal to $-1.5 \pm 0.2$. Interestingly, this value is consistent with that averaged over a sample of massive early-type galaxies (\citealt{gri12}).

The results of our strong lensing plus stellar population analysis (assuming a Salpeter IMF) can equivalently be interpreted as a very high projected dark over total mass fraction of about 80\% within a cylinder of radius given by the effective radius of the BG galaxy. Taken at face value, this dark matter fraction appears exceptional when compared to central dark matter fractions estimated for low- and intermediate-redshift massive galaxies, either from strong lensing analysis or dynamical modelling. In particular, \citet{gri09} (see also \citealt{aug10}) from a combined strong lensing and stellar population analysis of 57 early-type galaxies from the SLACS survey, and assuming a Chabrier IMF, derive dark over total mass fractions, projected within approximately half the values of the galaxy effective radii, that vary between 0.4 and 0.7 over a stellar mass range of $10^{10.5}$--$10^{12}M_\odot$. Similar central dark matter fractions are found in five higher-redshift lenses with an analogous analysis performed by \citet{mor11}. Stellar population analyses combined with dynamical mass estimates of SDSS early-type galaxies also provide central projected dark over total mass fractions between 0.2 and 0.7, over a similar mass range and assuming a Chabrier IMF (\citealt{gal06}; \citealt{gri10a}). These values are consistent with the typical central three-dimensional dark-matter fractions estimated by dynamical modelling by \citet{tor12}, but higher than those obtained by \citet{cap12b} with 3D JAM dynamical modelling of 260 early-type galaxies in the ATLAS3D survey and without any assumption on the universality of the IMF. Thus, also considering the different IMF assumptions, the central dark over total mass fraction of CSWA 31 is at the highest end (or larger) than those typically measured in early-type galaxies of similar mass.

We should however keep in mind on the one hand the very high stellar mass of the BG analysed here and on the other the high-density environment in which it resides. Both these points could make an important difference in the expected amount of dark matter. The amount of dark matter that we measure in CSWA 31 is likely the sum of the halo of the BG and of a more extended halo of the group, which would clearly provide more dark matter than that estimated for early-type galaxies in lower density environments (although we note that no significant environmental dependence is detected by \citealt{tor12}). \citet{gas07} modelled the mass profiles of 16 galaxy groups with X-ray data and they found that a non-negligible stellar mass component is required in only half of their sample. For these systems the central dark over total mass fractions are on average approximately 60\% (from their figures 3, 4, 5, 6 and the effective radii in their table 2). Our system, with a BG about an order of magnitude more massive than those in \citet{gas07}, has a dark matter fraction higher than their average (interestingly, in \citealt{gas07}, the central dark matter fraction reaches 90\% only in the system with the least massive BG). A central dark matter fraction similar to what we derive in CSWA 31 is instead found in the fossil group RXC2315.7$-$0222 (\citealt{dem10}), where we infer a value of 86\% according to their NFW+stars fit in figure~6 and assuming an effective radius of 16~kpc (S. Zibetti, private communication).

Next, we use the SDSS spectrum shown in Fig. \ref{fig4} to estimate the stellar velocity
dispersion $\sigma_{*}$ of the BG. The signal-to-noise ratio per pixel of the spectrum
is relatively low and varies approximately from 3 at 6500 {\AA} to 5 at
8000 {\AA}. We use the pixel-fitting method of \citet{cappellari04} by
first shifting the spectrum to the rest-frame from the measured
redshift $z=0.683$ and compare the rest-frame spectrum at wavelengths
3820-5290~{\AA} with template stellar spectra from the MILES library
\citep{sanchez-blazquez06,falcon-barroso11}. The resulting stellar velocity
dispersion value is remarkably large $(450\pm80)$~km~s$^{-1}$ with a best-fit that gives
a value of reduced $\chi^2$ equal to 1.03. We caution that this last result might change if a higher signal-to-noise spectrum were available, but, we also notice that such a high stellar velocity dispersion is consistent with an extrapolation of the $\sigma$-$M_{*}$ relation observed in massive early-type lens galaxies (e.g., \citealt{aug10}). We remark that according to the estimated values of $M_{*}$ and $\sigma_{*}$ the BG might be among the most massive central galaxies identified so far and offer an interesting case for testing galaxy formation physics at the extreme end of the galaxy mass function (e.g., \citealt{mar12}; \citealt{tho12}). Moreover, following the prescriptions detailed in \cite{agn13}, i.e. assuming a power-law, spherical, total density profile for the deflector and using the virial theorem and the ray-tracing equation, we find that the combination of the dynamics and lensing observables provides an estimate of $\gamma_{\mathrm{T}}$ of $-1.65 \pm 0.05$, almost completely insensitive to any orbital anisotropy assumption and projection effect. This result is consistent with the values presented in Sect.~3 and obtained from strong lensing only. An unambiguous confirmation of the existence at intermediate redshifts of objects with values of $\sigma_{*}$ of approximately 500 km s$^{-1}$ would open some interesting cosmological questions about the star formation history of massive early-type galaxies and the value of the inner slope of their dark matter haloes (e.g., \citealt{loe03}). 

\subsection{CSWA~31 as a possible fossil system}

The stellar component of the CSWA~31 core is heavily dominated by that
of the BG, with a magnitude difference with respect to the second
brightest galaxy, G3, of $\Delta m_{12} = 2.1 \pm 0.2$ and $1.9 \pm
0.1$~mag in the observer-frame $r$- and $i$-band, respectively. As
such, CSWA~31 is optically reminiscent of ``fossil'' groups/clusters,
defined as systems with $\Delta m_{12} \ge 2.0$ in rest-frame $R$
within a projected radius of $0.5 \, R_{200}$, and with an X-ray
luminosity from intergalactic gas of $L_{\rm X} >
10^{42}$~erg~s$^{-1}$ (\citealt{jon03}).

To test whether CSWA~31 can in fact be classified as a fossil, we
first note that our inferred upper limit on $L_{\rm X}$ within
$R=150$~kpc is two orders of magnitude above the value of the
\citet{jon03} criterion. In the absence of deeper X-ray data, we hence cannot exclude that the system is consistent with this fossil
criterion. Moreover, the $L_{\rm X}$ criterion was originally imposed
in order to ensure the presence of a group-scale dark matter halo, but
this is clearly already established from our lensing analysis.  The
second requirement, $\Delta m_{12} \ge 2.0$ within $0.5 \, R_{200}$, can
be tested with the aid of our extrapolation of the total mass profile
from Section~4.2. This suggests a value of $0.5 \, R_{200} = 0.7$--1.5 Mpc
for a plausible range in NFW concentration parameters of $c_{200} =
3$--10 (with lower $c$ corresponding to higher $R_{200}$).
We therefore searched for potential group members within $R \le
1.5$~Mpc from the BG in projection, requiring them to be brighter than
G3 and having a photometric redshift consistent with the spectroscopic
value of the BG. A search within SDSS $i$-band data (which most
closely match rest-frame $R$ at these redshifts) returned two such
objects, at $z_{\mathrm{ph}}=0.63 \pm 0.08$ and $0.73 \pm 0.11$, with projected
radii $R = 760$--1040~kpc and for which $\Delta m_{12}=1.6$--1.8~mag.
Under the assumption of an NFW mass profile, these galaxies would
reside within $0.5 \, R_{200}$ for $c < 10$ and $< 5$, respectively, and
so would violate the fossil criterion if belonging to
CSWA~31. However, given the large uncertainties associated with
extrapolating the mass profile out to such large radii from the BG centre,
it is entirely possible that these two galaxies are located well
beyond the true value of $0.5 \, R_{200}$. Verification of the fossil
status of CSWA~31 would thus require a more reliable assessment of the
value of $R_{200}$ from, e.g., weak lensing or X-ray data, along with
spectroscopic redshift estimates of these two galaxies to assess their
possible group/cluster membership.

With the available data, we can only identify CSWA~31 as a {\em
  candidate} fossil system. If confirmed, it would be among the most
distant and massive fossils explored in detail so far
\citep{ulm05,kho06}, and represent only the second such system studied
with lensing methods \citep{sch10}. Deep X-ray data would be required
to test whether CSWA~31 meets the fossil X-ray criterion and is indeed
a dynamically evolved system with an undisturbed hot gas morphology,
as generally anticipated for fossils. It is also relevant to note that
the large values of $\Delta m_{12}$ in these systems are commonly
explained in terms of dynamical friction having caused all central
intermediate-mass galaxies to be cannibalized by the BG. However, for
galaxies on circular orbits, the timescale for this to occur in
CSWA~31 would almost certainly exceed the age of the universe at $z
\approx 0.7$ \citep{don05}. Confirmation of the presence of such a
massive fossil already at these redshifts would therefore impose
valuable constraints on the galaxy infall geometry and general
formation history of fossils.

\section{Conclusions}

We have studied the gravitational lensing system CSWA 31 in which a bright early-type galaxy at redshift 0.683 (with several nearby, smaller companions) acts as primary lens on a background source. The observed four lensed images of this source provide a large Einstein radius of approximately 70 kpc. We have modelled available photometric and spectroscopic data, in the optical and X-ray wavelength ranges, to explore the deflector total and baryonic mass components, through a joint strong gravitational lensing, composite stellar population and X-ray emission analysis. The main results of this work can be summarized as follows:

\begin{itemize}

\item[$\bullet$] From the detected emission lines in our new VLT/X-shooter data, we derive a source redshift of 1.4870.

\item[$\bullet$] For the deflector, we measure a total mass value projected within the Einstein radius of $(40 \pm 1)\times 10^{12}$ M$_{\odot}$ and an inner three-dimensional logarithmic slope of the total mass density of $-1.7 \pm 0.2$.

\item[$\bullet$] For the brightest galaxy, we obtain very high values of stellar mass and velocity dispersion of, respectively, $(3 \pm 1)$~$\times$~$10^{12}$~M$_{\odot}$ (assuming a Salpeter stellar IMF) and $(450 \pm 80$) km s$^{-1}$ (inside the SDSS 3\arcsec-diameter fibre).

\item[$\bullet$] We find an upper limit of approximately $3 \times 10^{12}$~M$_\odot$ within $R=150$~kpc for the hot gas mass component of the deflector.

\item[$\bullet$] We measure low projected stellar-to-total mass fractions of 20\% and 3--6\% within apertures equal to, respectively, 25 kpc (i.e., the effective radius of the brightest galaxy) and 100 kpc.

\end{itemize}

We remark that atypical strong lensing systems as CSWA 31 are particularly useful to test our understanding of the baryonic and dark-matter mass assembly history of cosmological structures, as imprinted in the mass profile, in the currently accepted $\Lambda$CDM scenario. The high stellar mass value of the brightest galaxy and the large magnitude gap between this and the other lens galaxies might classify CSWA 31 as the most distant fossil candidate studied in detail so far. Improved data on the multiple image systems, on the possible hot gas X-ray emission, and on the central stellar velocity of the brightest galaxy of CSWA 31 would allow to combine effectively the different mass diagnostics in order to measure more precisely the inner steepness and mass of the dark matter halo associated with the deflector. These results would offer invaluable information to be compared against cosmological simulations.

\section*{Acknowledgments}

The Dark Cosmology Centre is funded by the DNRF. L.C. acknowledges the support of the EU under a Marie Curie Intra-European Fellowship, contract PIEF-GA-2010-274117. A.G. acknowledges funding from the European Union Seventh Framework Programme (FP7/2007-2013) under grant agreement n$^\circ$ 267251 “Astronomy Fellowships in Italy” (AstroFIt). We also acknowledge the use of data from the SDSS data base (http://www.sdss.org/). We thank K. Denney for helping in carrying out the observations and A. Agnello for useful discussions.


\label{lastpage}


\begin{thebibliography}{99}

\bibitem[Agnello, Auger \& Evans(2013)]{agn13} Agnello A., Auger M. W., Evans N. W., 2013, MNRAS, 429, 35
\bibitem[Auger et al.(2009)]{aug09} Auger M. W., Treu T., Bolton A. S., Gavazzi R., Koopmans L. V. E., Marshall P. J., Bundy K., Moustakas L. A., 2009, ApJ, 705, 1099
\bibitem[Auger et al.(2010)]{aug10} Auger M. W., Treu T., Bolton A. S., Gavazzi R., Koopmans L. V. E., Marshall P. J., Moustakas L. A., 2010, ApJ, 724, 511
\bibitem[Barnab\`e et al.(2011)]{bar11} Barnab\`e M., Czoske O., Koopmans L. V. E., Treu T., Bolton A. S., 2011, MNRAS, 415, 2215
\bibitem[Barnab\`e et al.(2012)]{bar12} Barnab\`e M. et al., 2012, MNRAS, 423, 1073
\bibitem[Bartelmann (2010)]{bar10} Bartelmann M., 2010, CQGra, 27, 233001
\bibitem[Bell \& de Jong(2001)]{BdJ01} Bell E.~F., de Jong R.~S., 2001, ApJ, 550, 212 
\bibitem[Belokurov et al.(2009)]{bel09} Belokurov V., Evans N. W., Hewett P. C., Moiseev A., McMahon R. G., Sanchez S. F., King L. J., 2009, MNRAS, 392, 104
\bibitem[Brewer et al.(2011)]{brewer11} Brewer B.~J., Lewis 
G.~F., Belokurov V., Irwin M.~J., Bridges T.~J., Evans N.~W., 2011, MNRAS, 412, 2521 
\bibitem[Cappellari \& Emsellem(2004)]{cappellari04} Cappellari M., Emsellem E., 2004, PASP, 116, 138 
\bibitem[Cappellari et al.(2012a)]{cap12} Cappellari M. et al., 2012a, Nature, 484, 485
\bibitem[Cappellari et al.(2012b)]{cap12b} Cappellari M. et al., 2012b, arXiv:1208.3522 
\bibitem[Chabrier(2003)]{chabrier03} Chabrier G., 2003, PASP, 115, 763 
\bibitem[Charlot \& Fall(2000)]{CF00} Charlot S., Fall S.~M., 2000, ApJ, 539, 718 
\bibitem[Christensen et al.(2012)]{christensen12} Christensen L. et al., 2012, MNRAS, 427, 1953
\bibitem[\protect\citeauthoryear{Clowe et al.}{2006}]{clow06} 
  Clowe D., Brada{\v c} M., Gonzalez A.~H., Markevitch M., Randall S.~W., 
  Jones C., Zaritsky D., 2006, ApJ, 648, 109 
\bibitem[Conroy \& van Dokkum(2012)]{con12} Conroy C., van Dokkum P., 2012, ApJ, 760, 61
\bibitem[D{\'e}mocl{\`e}s et al.(2010)]{dem10} D{\'e}mocl{\`e}s J., Pratt G.~W., Pierini D., Arnaud M., Zibetti S., D'Onghia E., 2010, A\&A, 517, 52 
\bibitem[D'Onghia et al.(2005)]{don05} D'Onghia E., Sommer-Larsen J., Romeo A. D., Burkert A., Pedersen K., Portinari L., Rasmussen, J., 2005, ApJ, 630, 109
\bibitem[Eichner, Seitz \& Bauer (2012)]{eic12} Eichner T., Seitz S., Bauer A., 2012, MNRAS, 427, 1918
\bibitem[Falc{\'o}n-Barroso et al.(2011)]{falcon-barroso11} Falc{\'o}n-Barroso J., S{\'a}nchez-Bl{\'a}zquez P., Vazdekis A., Ricciardelli E., Cardiel N., Cenarro A. J., Gorgas J., Peletier R. F., 2011, A\&A, 532, A95 
\bibitem[Gallazzi et al.(2005)]{gal05} Gallazzi A., Charlot S., Brinchmann J., White S.~D.~M., Tremonti C.~A., 2005, MNRAS, 362, 41
\bibitem[Gallazzi et al.(2006)]{gal06} Gallazzi A., Charlot S., Brinchmann J., White S.~D.~M., 2006, MNRAS, 370, 1106
\bibitem[Gastaldello et al.(2007)]{gas07} Gastaldello F., Buote D. A., Humphrey P. J., Zappacosta L., Bullock J. S., Brighenti F., Mathews W. G., 2007, ApJ, 669, 158
\bibitem[Grillo et al.(2008)]{gri08b} Grillo C. et al., 2008, A\&A, 486, 45
\bibitem[Grillo, Lombardi \& Bertin(2008)]{gri08c} Grillo C., Lombardi M., Bertin G. 2008, A\&A, 477, 397
\bibitem[Grillo et al.(2009)]{gri09} Grillo C., Gobat R., Lombardi M., Rosati P., 2009, A\&A, 501, 461
\bibitem[Grillo(2010)]{gri10a} Grillo C., 2010, ApJ, 722, 779
\bibitem[Grillo \& Gobat(2010)]{gri10b} Grillo C., Gobat R., 2010, MNRAS, 402, 67
\bibitem[Grillo et al.(2010)]{gri10c} Grillo C., Eichner T., Seitz S., Bender R., Lombardi M., Gobat R., Bauer A., 2010, ApJ, 710, 372
\bibitem[Grillo \& Christensen(2011)]{gri11} Grillo C., Christensen L., 2011, MNRAS, 418, 929
\bibitem[Grillo (2012)]{gri12} Grillo C., 2012, ApJ, 747, 15
\bibitem[Jones et al.(2003)]{jon03} Jones L. R., Ponman T. J., Horton A., Babul A., Ebeling H., Burke, D. J., 2003, MNRAS, 343, 627
\bibitem[Keeton(2001a)]{kee01a} Keeton C. R., 2001a, arXiv:astro-ph/0102340
\bibitem[Keeton(2001b)]{kee01b} Keeton C. R., 2001b, arXiv:astro-ph/0102341
\bibitem[\protect\citeauthoryear{Khosroshahi et al.}{2006}]{kho06} 
  Khosroshahi H.~G., Maughan B.~J., Ponman T.~J., Jones L.~R., 2006, MNRAS, 
  369, 1211 
\bibitem[Limousin et al.(2009)]{lim09} Limousin M. et al., 2009, A\&A, 502, 445
\bibitem[Loeb \& Peebles(2003)]{loe03} Loeb A., Peebles P.~J.~E., 2003, ApJ, 589, 29
\bibitem[Maraston et al.(2012)]{mar12} Maraston C. et al., 2012, arXiv:1207.6114
\bibitem[Modigliani et al.(2010)]{modigliani10} Modigliani A. et al., 2010, in Society of Photo-Optical Instrumentation
Engineers (SPIE) Conference Series, Vol. 7737, Society
of Photo-Optical Instrumentation Engineers (SPIE) Conference Series
\bibitem[More et al.(2011)]{mor11} More A., Jahnke K., More S., Gallazzi A., Bell E. F., Barden M., H{\"a}u{\ss}ler B., 2011, ApJ, 734, 69 
\bibitem[\protect\citeauthoryear{Navarro, Frenk \& White}
  {Navarro et al.}{1997}]{nava97} 
  Navarro J.~F., Frenk C.~S., White S.~D.~M., 1997, ApJ, 490, 493 
\bibitem[Newman et al.(2013)]{new12} Newman A. B., Treu T., Ellis R. S., Sand D. J., 2013, ApJ, 765, 25 
\bibitem[Salim et al.(2005)]{Salim05} Salim S. et al., 2005, ApJ, 619, 39 
\bibitem[Salpeter(1955)]{sal55} Salpeter E. E., 1955, ApJ, 121, 161
\bibitem[S{\'a}nchez-Bl{\'a}zquez et al.(2006)]{sanchez-blazquez06} 
S{\'a}nchez-Bl{\'a}zquez P. et al., 2006, MNRAS, 371, 703
\bibitem[\protect\citeauthoryear{Schirmer et al.}{2010}]{sch10} 
  Schirmer M., Suyu S., Schrabback T., Hildebrandt H., Erben T., Halkola A., 
  2010, A\&A, 514, A60 
\bibitem[Schwab, Bolton \& Rappaport(2010)]{schw10} Schwab J., Bolton A. S., Rappaport S. A., 2010, ApJ, 708, 750
\bibitem[Sonnenfeld et al.(2012)]{son12} Sonnenfeld A., Treu T., Gavazzi R., Marshall P. J., Auger M. W., Suyu S. H., Koopmans L. V. E., Bolton A. S., 2012, ApJ, 752, 163
\bibitem[Spiniello et al.(2011)]{spi11} Spiniello C., Koopmans L. V. E., Trager S. C., Czoske O., Treu T., 2011, MNRAS, 417, 3000
\bibitem[Suyu et al.(2010)]{suy10} Suyu S. H., Marshall P. J., Auger M. W., Hilbert S., Blandford R. D., Koopmans L. V. E., Fassnacht C. D., Treu T., 2010, ApJ, 711, 201
\bibitem[Suyu et al.(2013)]{suy13} Suyu S. H. et al., 2013, ApJ, 766, 70
\bibitem[Thomas et al.(2013)]{tho12} Thomas D. et al., 2013, MNRAS, 431, 1383
\bibitem[Tortora et al.(2012)]{tor12} Tortora C., La Barbera F., Napolitano N.~R., de Carvalho R.~R., Romanowsky A.~J., 2012, MNRAS, 425, 577
\bibitem[Treu (2010)]{tre10b} Treu T., 2010, ARA\&A, 48, 87
\bibitem[Treu et al.(2010)]{tre10} Treu T., Auger M. W., Koopmans L. V. E., Gavazzi R., Marshall P. J., Bolton A. S., 2010, ApJ, 709, 1195
\bibitem[\protect\citeauthoryear{Ulmer et al.}{2005}]{ulm05} Ulmer M.~P. et al., 2005, ApJ, 624, 124 
\bibitem[van Dokkum \& Conroy(2010)]{van10} van Dokkum P., Conroy C., 2010, Nature, 468, 940
\bibitem[Vegetti et al.(2010)]{veg10} Vegetti S., Koopmans L. V. E., Bolton A., Treu T., Gavazzi R., 2010, MNRAS, 408, 1969
\bibitem[Vegetti et al.(2012)]{veg12} Vegetti S., Lagattuta D. J., McKean J. P., Auger M. W., Fassnacht C. D., Koopmans L. V. E., 2012, Nature, 481, 341
\bibitem[Vernet et al.(2011)]{vernet11} Vernet J. et al., 2011, A\&A, 536, A105  
\bibitem[\protect\citeauthoryear{Vikhlinin et al.}{2006}]{vikh06} 
  Vikhlinin A., Kravtsov A., Forman W., Jones C., Markevitch M., Murray S.~S., 
  Van Speybroeck L., 2006, ApJ, 640, 691 
\bibitem[Zibetti, Charlot \& Rix(2009)]{zib09} Zibetti S., Charlot S., Rix H.-W., 2009, MNRAS, 400, 1181
\bibitem[Zitrin et al.(2012)]{zit12} Zitrin A. et al., 2012, ApJ, 749, 97
\end{thebibliography}
\end{document}